\documentclass[pre,aps,preprint,showpacs,preprintnumbers,amsmath,amssymb,secnumroman,eqsecnum]{revtex4}

\usepackage{graphicx}
\usepackage{dcolumn}
\usepackage{bm}

\newcommand{\beq}{\begin{equation}}
\newcommand{\eeq}{\end{equation}}
\newcommand{\beqa}{\begin{eqnarray}}
\newcommand{\eeqa}{\end{eqnarray}}
\newcommand{\vc}[1]{\mbox{\boldmath $#1$}}

\newcommand{\vol}[1]{{\bf #1}}

\begin{document}


\title{Stokesian swimmers and active particles}

\author{B. U. Felderhof}

 \email{ufelder@physik.rwth-aachen.de}
\affiliation{Institut f\"ur Theorie der Statistischen Physik \\ RWTH Aachen University\\
Templergraben 55\\52056 Aachen\\ Germany\\
}%

\date{\today}

\begin{abstract}
The net steady state flow pattern of a distorting sphere is studied in the framework of the bilinear theory of swimming at low Reynolds number. It is argued that the starting point of a theory of interacting active particles should be based on such a calculation, since any arbitrarily chosen steady state flow pattern is not necessarily the result of a swimming motion. Furthermore, it is stressed that as a rule the phase of stroke is relevant in hydrodynamic interactions, so that the net flow pattern must be used with caution.

\end{abstract}

\maketitle
\section{\label{I}Introduction}

The dynamics of swarms of active particles has been studied intensively in recent years \cite{1}-\cite{5}. Much of the work is based on the assumption that each particle of the swarm moves with a velocity determined by its own activity and by the local fluid flow velocity arising from the flow patterns of surrounding particles. The flow pattern of each particle is centered on that particle and is carried along with the particle velocity. Due to the perpetual change of particle positions this leads to an interesting many-body problem with complicated dynamics.

The assumptions in the theory, as used in practice, can be questioned on two counts. First, it is usually assumed that at each point in time the net steady state flow pattern of each particle is all that needs to be considered.  In fact the net flow pattern must be regarded as the time average over a period of the swimming or flying motion. On the fast time scale of the period there is an additional oscillating flow pattern. The phase of the oscillating pattern is important and affects the hydrodynamic interaction and hence the swimming velocities \cite{6}-\cite{8}. Second, a net steady state flow pattern is often assumed without derivation from a swimming motion on the fast time scale.

In the following we study the second assumption on the basis of low Reynolds number hydrodynamics \cite{9}. Therefore the fluid equations of motion are Stokes equations for a viscous incompressible fluid, and inertia effects are neglected. In Stokes hydrodynamics the flow at each point in space is determined instantaneously by the no-slip boundary condition on the surface of each of the particles.

We study a single distorting sphere and calculate the resulting net flow pattern to second order in the amplitude of stroke. It turns out that a commonly assumed active particle flow pattern, of so-called $B_1B_2$ type, cannot be realized as the result of the swimming motion of a distorting sphere. In particular this calls into question the calculation of the hydrodynamic interaction of two swimming micro-organisms for which the $B_1B_2$ model was first proposed \cite{10}.

We conclude that instead of assuming a particular net steady state flow pattern for an active particle it is preferable to consider a swimmer characterized by a combination of low order oscillating multipole moments and to calculate the corresponding net flow pattern. Several examples of such explicit calculations are presented. Each of the resulting net flow patterns can be used in the dynamics of swarms of active particles, though with the caveat that the phase of stroke may be relevant in hydrodynamic interactions.

\section{\label{II}Swimming sphere}

We consider a sphere of radius $a$ immersed in a viscous incompressible fluid of shear viscosity $\eta$. The fluid is of infinite extent and at rest at infinity. It is made to move as a result of shape deformations of the sphere, which change the undeformed sphere with surface $S_0$ into a body with surface $S(t)$ at time $t$. The fluid flow equations are formulated conveniently in the instantaneous rest frame of the body. It is assumed that in this frame the flow velocity $\vc{v}(\vc{r},t)$ and the pressure $p(\vc{r},t)$ satisfy the Stokes equations
of low Reynolds number hydrodynamics \cite{9}
\begin{equation}
\label{2.1}\eta\nabla^2\vc{v}-\nabla p=0,\qquad\nabla\cdot\vc{v}=0.
\end{equation}
The flow velocity is assumed to satisfy the no-slip condition at the surface $S(t)$. A point on the surface $S_0$ of the undeformed sphere is denoted by $\vc{s}$, and the corresponding point on the surface $S(t)$ is denoted by $\vc{s}+\vc{\xi}(\vc{s},t)$, with displacement vector $\vc{\xi}(\vc{s},t)$. The no-slip condition reads \cite{11}
\begin{equation}
\label{2.2}\vc{v}(\vc{s}+\vc{\xi}(\vc{s},t))=\frac{\partial\vc{\xi}(\vc{s},t)}{\partial t}.
\end{equation}

We place the origin of a Cartesian system of coordinates at the center of the sphere $S_0$. By definition
\begin{equation}
\label{2.3}\int_{S_0}\vc{\xi}\;dS=0.
\end{equation}
We also exclude the radial displacement corresponding to uniform expansion of the sphere.
We assume for simplicity that the displacement is axially symmetric and choose the axis of symmetry as $z$ axis. As a consequence the flow velocity and pressure are also axially symmetric, and the body acquires a translational velocity $\vc{U}(t)=U(t)\vc{e}_z$ in the direction of the $z$ axis, but no rotational velocity.

In spherical coordinates $(r,\theta,\varphi)$ the flow velocity can be expanded in terms of a set of fundamental solutions of the Stokes equations (2.1),
\begin{equation}
\label{2.4}\vc{v}(\vc{r},t)=-U(t)\vc{e}_z+\sum^\infty_{l=1}m_l(t)\vc{u}_l(r,\theta)+\sum^\infty_{l=2}k_l(t)\vc{v}_l(r,\theta),
\end{equation}
with \cite{12}
\begin{eqnarray}
\label{2.5}\vc{u}_l(r,\theta)&=&\bigg(\frac{a}{r}\bigg)^{l+2}\big[(l+1)P_l(\cos\theta)\vc{e}_r+P^1_l(\cos\theta)\vc{e}_\theta\big],\nonumber\\
\vc{v}_l(r,\theta)&=&\bigg(\frac{a}{r}\bigg)^l\big[(l+1)P_l(\cos\theta)\vc{e}_r+\frac{l-2}{l}P^1_l(\cos\theta)\vc{e}_\theta\big],
\end{eqnarray}
with Legendre polynomials $P_l(\cos\theta)$ and associated Legendre functions $P^1_l(\cos\theta)$ in the notation of Edmonds \cite{13}. In the second sum in Eq. (2.4) the term $\vc{v}_1(r,\theta)$ is missing on account of the requirement that the body exert no net force on the fluid. It follows from the expansion Eq. (2.4) that the translational velocity $U(t)$ may be calculated from the identity
\begin{equation}
\label{2.6}\vc{U}(t)=-\frac{1}{4\pi b^2}\int_{r=b}\vc{v}(\vc{r},t)\;dS,
\end{equation}
where the integral is over any large sphere centered at the origin and enclosing the body completely. The pressure corresponding to Eq. (2.4) is
\begin{equation}
\label{2.7}p(\vc{r},t)=p_0+2\eta\sum^\infty_{l=2}(2l-1)k_l(t)\frac{a^l}{r^{l+1}}P_l(\cos\theta),
\end{equation}
where $p_0$ is the ambient pressure at infinity.

Provided the sums in Eq. (2.4) converge we can use the expression also for $r=a$ and write
\begin{equation}
\label{2.8}\vc{v}(\vc{r},t)|_{r=a}=\sum^\infty_{l=1}A_l(t)P_l(\cos\theta)\vc{e}_r+\sum^\infty_{l=1}B_l(t)\frac{2}{l(l+1)}P^1_l(\cos\theta)\vc{e}_\theta,
\end{equation}
which defines the coefficients $\{A_l,B_l\}$ of Lighthill \cite{14} and Blake \cite{15}. By comparing Eq. (2.8) with Eq. (2.4) we find the relations
\begin{eqnarray}
\label{2.9}A_1&=&2m_1-U,\qquad B_1=m_1+U,\nonumber\\
A_l&=&(l+1)m_l+(l+1)k_l,\qquad B_l=\frac{1}{2}l(l+1)m_l+\frac{1}{2}(l-2)(l+1)k_l,\qquad (l>1).
\end{eqnarray}
The displacement may be written analogously to Eq. (2.4) as
\begin{equation}
\label{2.10}\vc{\xi}(\vc{s},t)=\sum^\infty_{l=1}M_l(t)\vc{u}_l(a,\theta)+\sum^\infty_{l=2}K_l(t)\vc{v}_l(a,\theta),
\end{equation}
where the term with $\vc{v}_1(a,\theta)$ is missing on account of Eq. (2.3).

We note in particular
 \begin{eqnarray}
\label{2.11}\vc{u}_1(r,\theta)&=&\bigg(\frac{a}{r}\bigg)^3\;[2\cos\theta\;\vc{e}_r+\sin\theta\;\vc{e}_\theta]=\frac{a^3}{r^3}(-\vc{I}+3\vc{e}_r\vc{e}_r)\cdot\vc{e}_z,\nonumber\\
\vc{v}_1(r,\theta)&=&\frac{a}{r}\;[2\cos\theta\;\vc{e}_r-\sin\theta\;\vc{e}_\theta]=\frac{a}{r}(\vc{I}+\vc{e}_r\vc{e}_r)\cdot\vc{e}_z,\nonumber\\
\vc{u}_2(r,\theta)&=&\frac{3a^4}{4r^4}\;[(1+3\cos2\theta)\;\vc{e}_r+2\sin2\theta\;\vc{e}_\theta],\nonumber\\
\vc{v}_2(r,\theta)&=&\frac{3a^2}{4r^2}\;(1+3\cos2\theta)\vc{e}_r,
\end{eqnarray}
where $\vc{I}$ is the unit tensor. The field $\vc{u}_1$ is identical to an electrostatic dipole field, the field $\vc{v}_1$ is an Oseen monopole flow, the field $\vc{u}_2$ is identical to an electrostatic quadrupole field, and the field $\vc{v}_2$ is a hydrodynamic stresslet or Oseen dipole flow.

It is convenient to expand the flow velocity $\vc{v}$ and the pressure $p$ in powers of the displacement $\vc{\xi}$ as \cite{11}
\begin{equation}
\label{2.12}\vc{v}=\vc{v}^{(1)}+\vc{v}^{(2)}+...,\qquad p=p^{(1)}+p^{(2)}+....
\end{equation}
By expanding the no-slip boundary condition Eq. (2.2) we find that the velocity at the undisplaced surface is given by \cite{11}
 \begin{equation}
\label{2.13}\vc{u}^{(1)}_S=\vc{v}^{(1)}|_{r=a}=\frac{\partial\vc{\xi}}{\partial t},\qquad\vc{u}^{(2)}_S=\vc{v}^{(2)}|_{r=a}=-\vc{\xi}\cdot\nabla\vc{v}^{(1)}|_{r=a}.
\end{equation}
The translational velocity $\vc{U}(t)$ has the corresponding expansion
\begin{equation}
\label{2.14}\vc{U}=\vc{U}^{(2)}+\vc{U}^{(3)}+....
\end{equation}
Here the first order term is missing on account of Eq. (2.3). From Eq. (2.9) this implies $A_1^{(1)}=2m^{(1)}_1$ and $B_1^{(1)}=m_1^{(1)}$. The second order term in the velocity is given by
\begin{equation}
\label{2.15}\vc{U}^{(2)}=-\frac{1}{4\pi a^2}\int_{S_0}\vc{u}^{(2)}_S\;dS.
\end{equation}

For periodic displacements with period $T=2\pi/\omega$ we put
\begin{eqnarray}
\label{2.16}M_l(t)=a(\mu_{ls}\cos\omega t-\mu_{lc}\sin\omega t),\nonumber\\
K_l(t)=a(\kappa_{ls}\cos\omega t-\kappa_{lc}\sin\omega t),
\end{eqnarray}
with dimensionless coefficients $\mu_{ls},\mu_{lc},\kappa_{ls},\kappa_{lc}$. Then we have from Eq. (2.13) for the first order flow velocity
\begin{equation}
\label{2.17}\vc{v}^{(1)}(\vc{r},t)=-a\omega\bigg[\sum^\infty_{l=1}\mu_l(t)\vc{u}_l(r,\theta)+\sum^\infty_{l=2}\kappa_l(t)\vc{v}_l(r,\theta)\bigg],
\end{equation}
with multipole coefficients
\begin{eqnarray}
\label{2.18}\mu_l(t)=\mu_{lc}\cos\omega t+\mu_{ls}\sin\omega t,\nonumber\\
\kappa_l(t)=\kappa_{lc}\cos\omega t+\kappa_{ls}\sin\omega t,
\end{eqnarray}
in the notation of Felderhof and Jones \cite{12}. The velocity $U^{(2)}(t)$ and the rate of dissipation $\mathcal{D}^{(2)}(t)$ to second order in the displacement can be expressed as bilinear expressions \cite{12} in terms of the coefficients $\{\mu_{ls},\mu_{lc},\kappa_{ls},\kappa_{lc}\}$.

From Eqs. (2.4) and (2.17) we find for the first order moments
\begin{equation}
\label{2.19}m^{(1)}_l(t)=-a\omega\mu_l(t),\qquad k^{(1)}_l(t)=-a\omega\kappa_l(t).
\end{equation}
In particular, from Eq. (2.9)
\begin{equation}
\label{2.20}A_1^{(1)}(t)=-2a\omega\mu_1(t),\qquad B^{(1)}_1(t)=-a\omega\mu_1(t),
\end{equation}
since $U^{(1)}(t)=0$.

\section{\label{III}Net flow pattern}

In this section we consider a periodic swimmer with first order flow velocity given by Eq. (2.17), and calculate the mean second order flow pattern. The mean is calculated as the time average over a single period $T=2\pi/\omega$. Thus we consider
\begin{equation}
\label{3.1}\overline{\vc{v}^{(2)}(\vc{r})}=\frac{1}{T}\int^T_0\vc{v}^{(2)}(\vc{r},t)\;dt
\end{equation}
corresponding to some stroke or displacement $\vc{\xi}(\vc{s},t)$. Here the flow pattern $\vc{v}^{(2)}(\vc{r},t)$ is defined in the instantaneous rest frame at time $t$ and is the solution of the Stokes equations (2.1) which tends to $-U^{(2)}(t)\vc{e}_z$ at infinity and has boundary value at $r=a$ given by Eq. (2.13).

The second order time-averaged flow pattern may be expanded as
\begin{equation}
\label{3.2}\overline{\vc{v}^{(2)}(\vc{r})}=-\overline{U^{(2)}}\vc{e}_z+\sum^\infty_{l=1}\overline{m_l^{(2)}}\vc{u}_l(r,\theta)+\sum^\infty_{l=2}\overline{k_l^{(2)}}\vc{v}_l(r,\theta),
\end{equation}
corresponding to Eq. (2.4). At the surface the mean second order flow pattern is given by
\begin{equation}
\label{3.3}\overline{\vc{v}^{(2)}(\vc{r})|_{r=a}}=\overline{\vc{u}_S^{(2)}(\vc{s})}=-\overline{\vc{\xi}\cdot\nabla\vc{v}^{(1)}|_{r=a}}.
\end{equation}
The right hand side can be calculated for a given stroke $\vc{\xi}(\vc{s},t)$.
The flow pattern $\overline{\vc{v}^{(2)}(\vc{r})}$ in Eq. (3.2) tends to $-\overline {U^{(2)}}\vc{e}_z$ at infinity with value given by Eq. (2.15). We denote the corresponding coefficients given by the analogue of Eq. (2.8) as $\{A^{\prime}_l,B^{\prime}_l\}=\{\overline{A_l^{(2)}},\overline{B_l^{(2)}}\}$. These may be calculated from $\overline{\vc{u}_S^{(2)}(\vc{s})}$ by using the orthonormality relations of the Legendre functions \cite{13}. It may be checked that $A^{\prime}_1$ and $B^{\prime}_1$ satisfy the relation
 \begin{equation}
\label{3.4}\overline {U^{(2)}}=\frac{1}{3}(2B^{\prime}_1-A^{\prime}_1).
\end{equation}

We define the corresponding net flow pattern as
 \begin{equation}
\label{3.5}\vc{v}^\prime(\vc{r})=\overline{\vc{v}^{(2)}(\vc{r})}+\overline {U^{(2)}}\vc{e}_z.
\end{equation}
This tends to zero at infinity at least as fast as $1/r^2$ and can be identified with the flow pattern of an active particle. Conversely the question arises as to whether a chosen steady state flow pattern can be identified with the net flow $\vc{v}^\prime(\vc{r})$ of a periodic swimmer.
For example, can we find a stroke $\vc{\xi}(\vc{s},t)$ for which all coefficients $\{A^{\prime}_l\}$ vanish and only $B^{\prime}_1,B^{\prime}_2$ differ from zero?

\section{\label{IV}Simple swimmers as active particles}

In this section we consider some simple swimmers with strokes characterized by combinations of low order multipole moments. The analysis suggests the flow pattern of corresponding active particles.

The calculation of the time-average $\overline{\vc{u}_S^{(2)}(\vc{s})}$ in Eq. (3.3) is performed most easily by using complex notation
\begin{eqnarray}
\label{4.1}\mu^c_l&=&\mu_{lc}+i\mu_{ls},\qquad\mu_l(t)=\mu^c_le^{-i\omega t},\nonumber\\
\kappa^c_l&=&\kappa_{lc}+i\kappa_{ls},\qquad\kappa_l(t)=\kappa^c_le^{-i\omega t},
\end{eqnarray}
with the identity
\begin{equation}
\label{4.2}\overline{\vc{u}_S^{(2)}(\vc{s})}=-\frac{1}{2}\mathrm{Re}\;\vc{\xi}^*\cdot\nabla\vc{v}^{(1)}|_{r=a},
\end{equation}
with complex $\vc{v}^{(1)}$ given by Eq. (2.17) with complex coefficients $\{\mu_l(t),\kappa_l(t)\}$, and complex $\vc{\xi}$ given by
\begin{equation}
\label{4.3}\vc{\xi}=-ia\bigg[\sum^\infty_{l=1}\mu_l^c\vc{u}_l(a,\theta)+\sum^\infty_{l=2}\kappa_l^c\vc{v}_l(a,\theta)\bigg]e^{-i\omega t}.
\end{equation}
We have checked the expressions for the $\{A^{\prime}_l,B^{\prime}_l\}$-coefficients given below by a separate calculation of the coefficients $\{A_l^{(2)}(t),B_l^{(2)}(t)\}$ and a subsequent time-average.

We consider first a simple swimmer with only $\mu^c_1,\kappa^c_2,\mu^c_2$ different from zero, corresponding to the superposition of a potential dipole field, an Oseen dipole, and a potential quadrupole field. The mean swimming velocity is derived from Eq. (2.15) as
\begin{equation}
\label{4.4}\overline{U^{(2)}}=\frac{3}{5}\;a\omega(\mu_{1c}\kappa_{2s}-\mu_{1s}\kappa_{2c}+5\mu_{1c}\mu_{2s}-5\mu_{1s}\mu_{2c}),
\end{equation}
in agreement with Eq. (7.8) of Ref. 12.
From Eq. (3.5) we find for the $A^{\prime},B^{\prime}$-coefficients
\begin{eqnarray}
\label{4.5}A^{\prime}_1&=&=\frac{3}{5}\;a\omega(\mu_{1c}\kappa_{2s}-\mu_{1s}\kappa_{2c}+3\mu_{1c}\mu_{2s}-3\mu_{1s}\mu_{2c}),\nonumber\\
B^{\prime}_1&=&\frac{3}{2}\overline{U^{(2)}}+\frac{1}{2}A^{\prime}_1,\nonumber\\
A^{\prime}_2&=&\frac{2}{5}B^{\prime}_2=-\frac{9}{7}\;a\omega(\mu_{2c}\kappa_{2s}-\mu_{2s}\kappa_{2c}),\nonumber\\
A^{\prime}_3&=&10\overline{U^{(2)}}-16A^{\prime}_1,\qquad B^{\prime}_3=21\overline{U^{(2)}}-33A^{\prime}_1,\nonumber\\
A^{\prime}_4&=&\frac{2}{5}B^{\prime}_4=6A^{\prime}_2.
\end{eqnarray}
The coefficients for $l>4$ vanish. The corresponding multipole moments $\{m_l^{\prime},\;k_l^{\prime}\}$ are calculated from the inverse of Eq. (2.9). The lowest order moments are
 \begin{equation}
\label{4.6}m_1^\prime=\frac{1}{2}(U^{(2)}+A^\prime_1),\qquad k^\prime_2=\frac{-1}{5}B^\prime_2,\qquad m^\prime_2=\frac{1}{3}B^\prime_2.
\end{equation}
Squirming at $l=2$ with $\mu^c_2=-\kappa^c_2$ implies $A^{\prime}_2=0,\;B^{\prime}_2=0$ and $A^\prime_1=\frac{1}{2}\overline{U^{(2)}}$. In the notation of Drescher et al. \cite{16} the source doublet strength is $A_{sd}=3m_1^{\prime}a^3$ and the stresslet strength is $A_{str}=\frac{3}{2}k_2^{\prime}a^2$. These authors measured the values for swimming Volvox, but they did not find a contribution from the quadrupole $m^{\prime}_2$, or higher order multipoles.

For the mean rate of dissipation or power we find from Eq. (7.15) of Ref. 12
 \begin{equation}
\label{4.7}\overline{\mathcal{D}^{(2)}}=8\pi\eta\omega^2a^3\frac{3}{20}\big[10(\mu_{1c}^2+\mu_{1s}^2)+9(\kappa^2_{2c}+\kappa^2_{2s})
+20(\mu_{2c}^2+\mu_{2s}^2)+24(\kappa_{2c}\mu_{2c}+\kappa_{2s}\mu_{2s})\big].
\end{equation}
The calculations in Eqs. (4.4) and (4.7) are performed most easily by using the expressions given in Eqs. (7.11) and (7.17) of Ref. 12.
From Eqs. (4.4) and (4.7) we derive an expression for the swimming efficiency defined by \cite{12}
 \begin{equation}
\label{4.8}E_T=4\eta\omega a^2\frac{|\overline{U_2}|}{\overline{\mathcal{D}_2}}.
\end{equation}
Without loss of generality we can choose the phase such that $\mu_{1s}=0$. We then find that the efficiency is maximal for
\begin{equation}
\label{4.9}\kappa_{2c}=0,\qquad\mu_{2c}=0,\qquad\kappa_{2s}=-\frac{4}{3}\sqrt{2}\;\mu_{1c},\qquad\mu_{2s}=\frac{11}{5\sqrt{2}}\;\mu_{1c},
\end{equation}
with value $E_T=5/(6\pi\sqrt{2})=0.188$. The net flow pattern $\vc{v}^\prime(\vc{r})$ can be calculated from the multipole coefficients $\{m_l^{\prime},\;k_l^{\prime}\}$ by using Eqs. (3.2) and (3.5). We note that the particular coefficients $A_2^{\prime}$ and $B_2^{\prime}$, as well as $A_4^{\prime}$ and $B_4^{\prime}$, vanish when Eq. (4.9) holds. It follows from Eq. (2.9) that then $m_2^{\prime},\;k_2^{\prime}$ and $m_4^{\prime},\;k_4^{\prime}$ also vanish. The net flow corresponds to a potential dipole of strength $m_1^{\prime}=(59\sqrt{2}/75)\mu_{1c}^2a\omega$ and equal multipoles at $l=3$ with $k_3^{\prime}=m_3^{\prime}=(459/472)m_1^{\prime}$. The measurements of Drescher et al. \cite{16} for Volvox do not correspond to the pattern for optimal swimming in the above sense, since they find $A_{str}$ to be different from zero.

In order to visualize the axisymmetric flow pattern it is useful to introduce a Stokes stream function $\psi$ via the relations \cite{17}
\begin{equation}
\label{4.10}v_r=\frac{1}{r^2\sin\theta}\frac{\partial\psi}{\partial\theta},\qquad v_\theta=\frac{-1}{r\sin\theta}\frac{\partial\psi}{\partial r}.
\end{equation}
The uniform flow $\vc{e}_z$ corresponds to the stream function
\begin{equation}
\label{4.11}\psi_z(\vc{r})=\frac{1}{2}\;r^2\sin^2\theta,
\end{equation}
and the flow patterns in Eq. (2.5) correspond to
\begin{equation}
\label{4.12}\psi_{ul}(r,\theta)=\frac{a^{l+2}}{lr^l}\;\sin\theta P^1_l(\cos\theta),\qquad
\psi_{vl}(r,\theta)=\frac{a^lr^2}{lr^{l}}\;\sin\theta P^1_l(\cos\theta).
\end{equation}

The streamlines of the flow are given by lines of constant $\psi$. In Fig. 1 we show the streamlines of the net flow $\vc{v}^\prime(\vc{r})$ calculated from $\overline{U^{(2)}}$ and the optimal moments of Eq. (4.9). In Fig. 2 we show the values of $v^{\prime }_r/\overline{U^{(2)}}$ and $v^{\prime }_\theta/\overline{U^{(2)}}$ at $r=a$ as functions of the polar angle $\theta$. In Fig. 3 we show the values of $\vc{v}^{\prime 2}/\overline{U^{(2)}}\;^2$ at $r=a$, $r=1.25a$, and $r=1.5a$ as functions of the polar angle $\theta$.

As a second example we consider a squirming swimmer characterized by coefficients $\mu^c_1=0,\;\kappa^c_2=-\mu^c_2,\; \kappa^c_3=-\mu^c_3$ with all higher order moments vanishing.
The mean swimming velocity is derived from Eq. (2.15), or from the expression in Eq. (7.11) of Ref. 12, as
\begin{equation}
\label{4.13}\overline{U^{(2)}}=\frac{48}{35}\;a\omega(\mu_{2c}\mu_{3s}-\mu_{2s}\mu_{3c}).
\end{equation}
For the mean rate of dissipation we find
 \begin{equation}
\label{4.14}\overline{\mathcal{D}^{(2)}}=8\pi\eta\omega^2a^3\frac{1}{12}\big[9(\mu_{2c}^2+\mu_{2s}^2)+8(\mu_{3c}^2+\mu_{3s}^2)\big].
\end{equation}

The calculation based on the analogue of Eq. (2.8) for the boundary value $\overline{\vc{u}^{(2)}_S(\vc{s})}$ of the flow pattern $\overline{\vc{v}^{(2)}(\vc{r})}$ shows that all $A^{\prime}$-coefficients vanish and it yields for the $B^{\prime}$-coefficients
\begin{eqnarray}
\label{4.15}B^{\prime}_1&=&\frac{3}{2}\;\overline{U^{(2)}},\qquad
B^{\prime}_2=0,\nonumber\\
B^{\prime}_3&=&\frac{7}{12}\;\overline{U^{(2)}},\qquad
B^{\prime}_4=0,\qquad B^{\prime}_5=-\frac{25}{12}\;\overline{U^{(2)}}.
\end{eqnarray}
The corresponding multipole moments are found from Eq. (2.9) as
\begin{eqnarray}
\label{4.16}m^\prime_1&=&\frac{1}{2}\;\overline{U^{(2)}},\qquad m^\prime_2=k^\prime_2=0,\nonumber\\
m^{\prime}_3&=&-k^{\prime}_3=\frac{7}{48}\;\overline{U^{(2)}},\qquad m^\prime_4=k^\prime_4=0,\nonumber\\
m^{\prime}_5&=&-k^{\prime}_5=-\frac{25}{72}\;\overline{U^{(2)}}.
\end{eqnarray}
The moments for $l>5$ vanish. The net flow pattern $\vc{v}^\prime(\vc{r})$ can be calculated from the multipole moments by using Eqs. (3.2) and (3.5).

Without loss of generality we can choose the phase such that $\mu_{2s}=0$. We then find that the efficiency $E_T$ is maximal for $\mu_{3c}=0,\;\mu_{3s}=\sqrt{9/8}\mu_{2c}$ with value $E_T=12\sqrt{2}/(35\pi)=0.154$.  In Fig. 4 we show the streamlines of the net flow $\vc{v}^\prime(\vc{r})$ calculated from $\overline{U^{(2)}}$ and the set of coefficients $\{A_l^{\prime},B_l^{\prime}\}$ given by Eq. (4.15) for the optimal moments. The net flow pattern at the surface $r=a$ is given by
 \begin{eqnarray}
\label{4.17}\vc{v}'|_{r=a}&=&\overline{\vc{u}_S^{(2)}(\vc{s})}+\overline{U^{(2)}}\vc{e}_z\nonumber\\
&=&\bigg[\cos\theta\;\vc{e}_r+\big[-\sin\theta+\frac{35}{64}(7+5\cos2\theta)\sin^3\theta\big]\;\vc{e}_\theta\bigg]\overline{U^{(2)}}.
\end{eqnarray}
 The radial component arises from the second term in the first line. In Fig. 5 we show the values of $v^{\prime }_r/\overline{U^{(2)}}$ and $v^{\prime }_\theta/\overline{U^{(2)}}$ at $r=a$ as functions of the polar angle $\theta$. In Fig. 6 we show the values of $\vc{v}^{\prime 2}/\overline{U^{(2)}}\;^2$ at $r=a$, $r=1.25a$, and $r=1.5a$ as functions of the polar angle $\theta$. It follows from Eq. (4.15) that the squirmer can be identified with a $B_1B_3B_5$-active particle with particular ratios of the coefficients.

 As a third example we consider a swimmer characterized by coefficients $\mu^c_1,\;\kappa^c_2,\;\kappa^c_3$, with all other moments vanishing. The mean swimming velocity is derived from Eq. (2.15), or from the expression in Eq. (7.11) of Ref. 12, as
\begin{equation}
\label{4.18}\overline{U^{(2)}}=\frac{3}{35}\;a\omega(7\mu_{1c}\kappa_{2s}-7\mu_{1s}\kappa_{2c}+6\kappa_{2c}\kappa_{3s}-6\kappa_{2s}\kappa_{3c}).
\end{equation}
For the mean rate of dissipation we find
 \begin{equation}
\label{4.19}\overline{\mathcal{D}^{(2)}}=8\pi\eta\omega^2a^3\frac{1}{420}\big[630(\mu_{1c}^2+\mu_{1s}^2)+567(\kappa_{2c}^2+\kappa_{2s}^2)+1180(\kappa_{3c}^2+\kappa_{3s}^2)\big].
\end{equation}

The calculation based on the analogue of Eq. (2.8) for the boundary value $\overline{\vc{u}^{(2)}_S(\vc{s})}$ of the flow pattern $\overline{\vc{v}^{(2)}(\vc{r})}$ yields for the non-vanishing $A^{\prime}$- and $B^{\prime}$-coefficients
\begin{eqnarray}
\label{4.20}A^{\prime}_1&=&\frac{1}{2}B^{\prime}_1=\overline{U^{(2)}},\nonumber\\
A^{\prime}_2&=&\frac{2}{5}B^{\prime}_2=\frac{20}{7}\;a\omega(\mu_{1c}\kappa_{3s}-\mu_{1s}\kappa_{3c}),\nonumber\\
A^{\prime}_3&=&\frac{1}{2}B^{\prime}_3=\frac{6}{5}\;a\omega(-3\mu_{1c}\kappa_{2s}+3\mu_{1s}\kappa_{2c}+\kappa_{2c}\kappa_{3s}-\kappa_{2s}\kappa_{3c}),\nonumber\\
A^{\prime}_4&=&\frac{2}{5}B^{\prime}_4=-A^\prime_2,\nonumber\\
A^{\prime}_5&=&\frac{1}{2}B^{\prime}_5=\frac{30}{7}\;a\omega(\kappa_{2c}\kappa_{3s}-\kappa_{2s}\kappa_{3c}).
\end{eqnarray}
Without loss of generality we can choose the phase such that $\mu_{1s}=0$. We then find that the efficiency is maximal for
\begin{equation}
\label{4.21}\kappa_{2c}=0,\qquad\kappa_{2s}=\frac{5}{3}\sqrt{\frac{230}{413}}\;\mu_{1c},\qquad\kappa_{3c}=-\frac{27}{59}\mu_{1c},\qquad\kappa_{3s}=0.
\end{equation}
with value $E_T=\sqrt{115/826}/(3\pi)=0.040$, showing that this type of swimming is rather less effective than for the first two examples. For the optimal swimmer the coefficients $A^{\prime}_2,\;B^{\prime}_2,\;A^{\prime}_4,\;B^{\prime}_4$ vanish.

Ghose and Adhikari \cite{18} have considered a similar swimmer with added potential flows with $m^c_2=-\frac{3}{5}k^c_2$ and $m^c_3=-\frac{5}{7}k^c_3$. In our analysis \cite{12} the $\{m_l\}$- and $\{k_l\}$-multipoles can have independent values. We recall that the above expressions are calculated from integrals bilinear in the first order flow velocity. The work of Ghose and Adhikari \cite{18} suggests that the time-dependent swimming velocity and flow pattern must be calculated to third order in the displacement $\vc{\xi}$ in order to agree with experimental observations of the swimming of Chlamydomonas \cite{19}. Delmotte et al. \cite{20} have studied the same swimmer in computer simulation.

We note that it follows from Eq. (2.7) that in all three examples the net flow pattern is accompanied by a steady pressure pattern $p^\prime(\vc{r})$. In the first and third examples the coefficients $A_2^{\prime}$ and $B_2^{\prime}$ are related by $A_2^{\prime}=\frac{2}{5}B_2^{\prime}$, and in the second example both coefficients vanish. It is not possible to design a swimmer with surface displacement (4.3) for which the steady state net second order flow pattern $\vc{v}^\prime(\vc{r})$ has all coefficients $\{A_l^{\prime}\}$ vanishing and only $B_1^{\prime},B_2^{\prime}$ of the $\{B_l^{\prime}\}$-coefficients nonvanishing.

\section{\label{V}Discussion}

The net steady state flow pattern of a periodically distorting sphere can be calculated for a chosen set of oscillating multipolar flow patterns determined by the stroke. As we have shown, such a calculation yields a net flow pattern characterized by a set of steady state multipole moments, as exemplified in Eqs. (4.6) and (4.16). In the study of hydrodynamic interactions between swimmers it is preferable to start from a particular set of oscillating multipole moments, rather than from a chosen steady state flow pattern. The net flow pattern must be used with caution since the hydrodynamic interactions may be affected by the relative phase of swimming strokes.

The explicit calculations of Sec. IV provide examples of simple multipolar swimmers. We suggest that the corresponding flow patterns be used in the study of the dynamics of swarms of active particles.
It may be preferable to use the stroke of optimum swimming efficiency within the chosen class of strokes.

\newpage

\section*{Figure captions}

\subsection*{Fig. 1}
Streamlines of the net flow $\vc{v}^\prime(\vc{r})$ for $\mu_{1c}=1,\;\mu_{1s}=0$, other moments given by Eq. (4.9), and vanishing higher order moments.

\subsection*{Fig. 2}
Plot of the components $v^{\prime }_r/\overline{U^{(2)}}$ (drawn curve) and $v^{\prime }_\theta/\overline{U^{(2)}}$ (dashed curve) of the net flow at $r=a$ as functions of $\theta$ for moments corresponding to Fig. 1.

\subsection*{Fig. 3}
Plot of $\vc{v}^{\prime 2}/\overline{U^{(2)}}\;^2$ at $r=a$ (drawn curve), $r=1.25a$ (long dashes), and $r=1.5a$ (short dashes) as functions of the polar angle $\theta$ for moments corresponding to Fig. 1.

\subsection*{Fig. 4}
Streamlines of the net flow $\vc{v}^\prime(\vc{r})$ for $\mu_{1c}=0,\;\mu_{1s}=0$, $\mu_{2c}=-\kappa_{2c}=1,\;\mu_{2s}=\kappa_{2s}=0$, $\mu_{3c}=\kappa_{3c}=0,\;\mu_{3s}=-\kappa_{3s}=\sqrt{9/8}$, and vanishing higher order moments.

\subsection*{Fig. 5}
Plot of the components $v^{\prime }_r/\overline{U^{(2)}}$ (drawn curve) and $v^{\prime }_\theta/\overline{U^{(2)}}$ (dashed curve) of the net flow at $r=a$ as functions of $\theta$ for moments corresponding to Fig. 4.

\subsection*{Fig. 6}
Plot of $\vc{v}^{\prime 2}/\overline{U^{(2)}}\;^2$ at $r=a$ (drawn curve), $r=1.25a$ (long dashes), and $r=1.5a$ (short dashes) as functions of the polar angle $\theta$ for moments corresponding to Fig. 4.

\newpage
\setlength{\unitlength}{1cm}
\begin{figure}
 \includegraphics{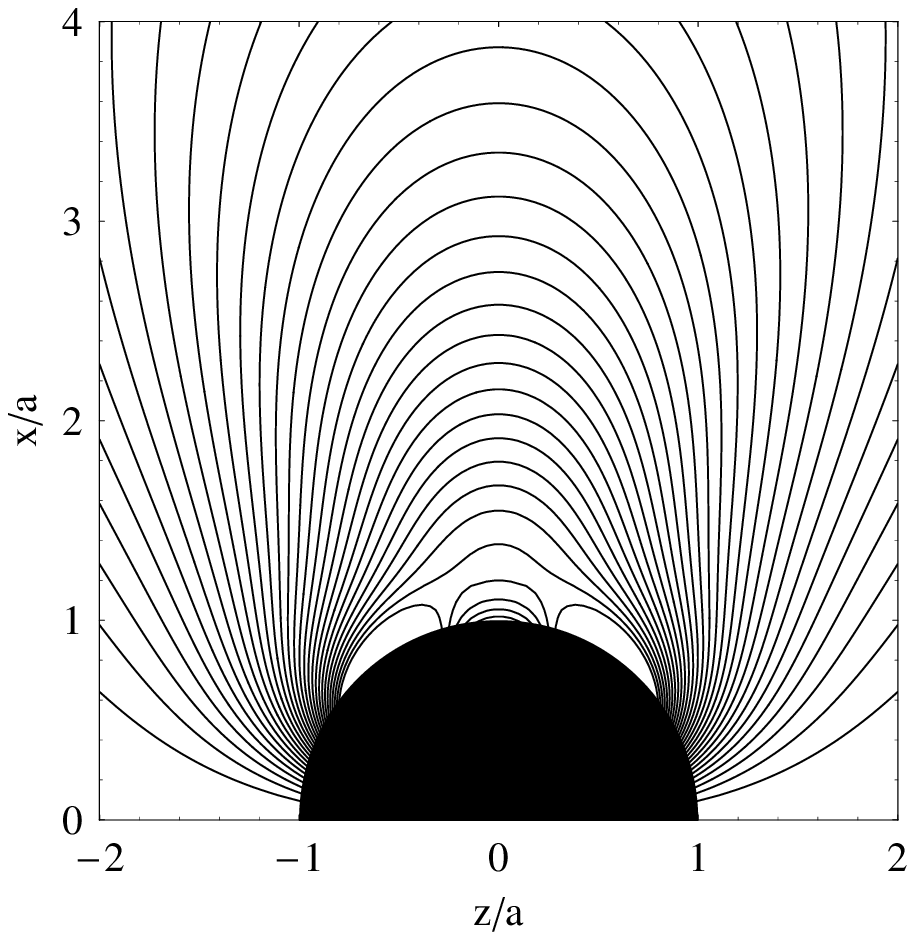}
   \put(-9.1,3.1){}
\put(-1.2,-.2){}
  \caption{}
\end{figure}
\newpage
\clearpage
\newpage
\setlength{\unitlength}{1cm}
\begin{figure}
 \includegraphics{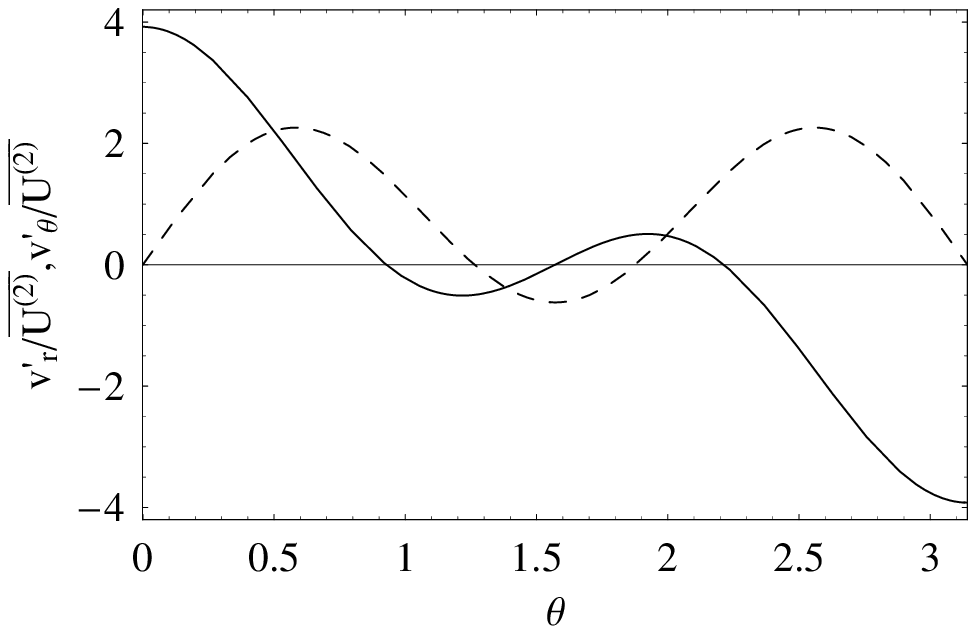}
   \put(-9.1,3.1){}
\put(-1.2,-.2){}
  \caption{}
\end{figure}
\newpage
\clearpage
\newpage
\setlength{\unitlength}{1cm}
\begin{figure}
 \includegraphics{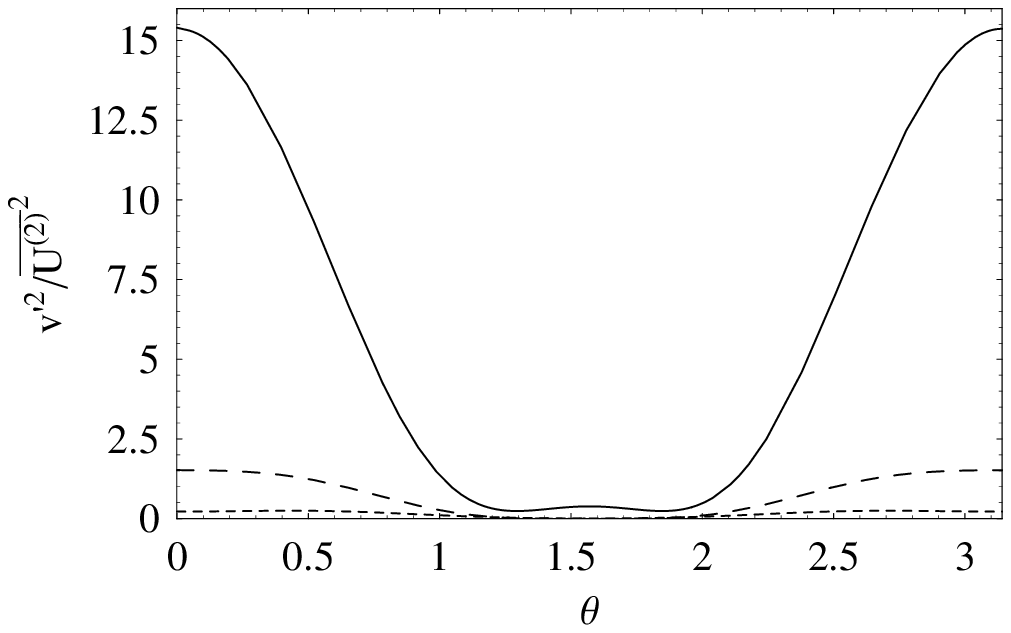}
   \put(-9.1,3.1){}
\put(-1.2,-.2){}
  \caption{}
\end{figure}
\newpage
\clearpage
\newpage
\setlength{\unitlength}{1cm}
\begin{figure}
 \includegraphics{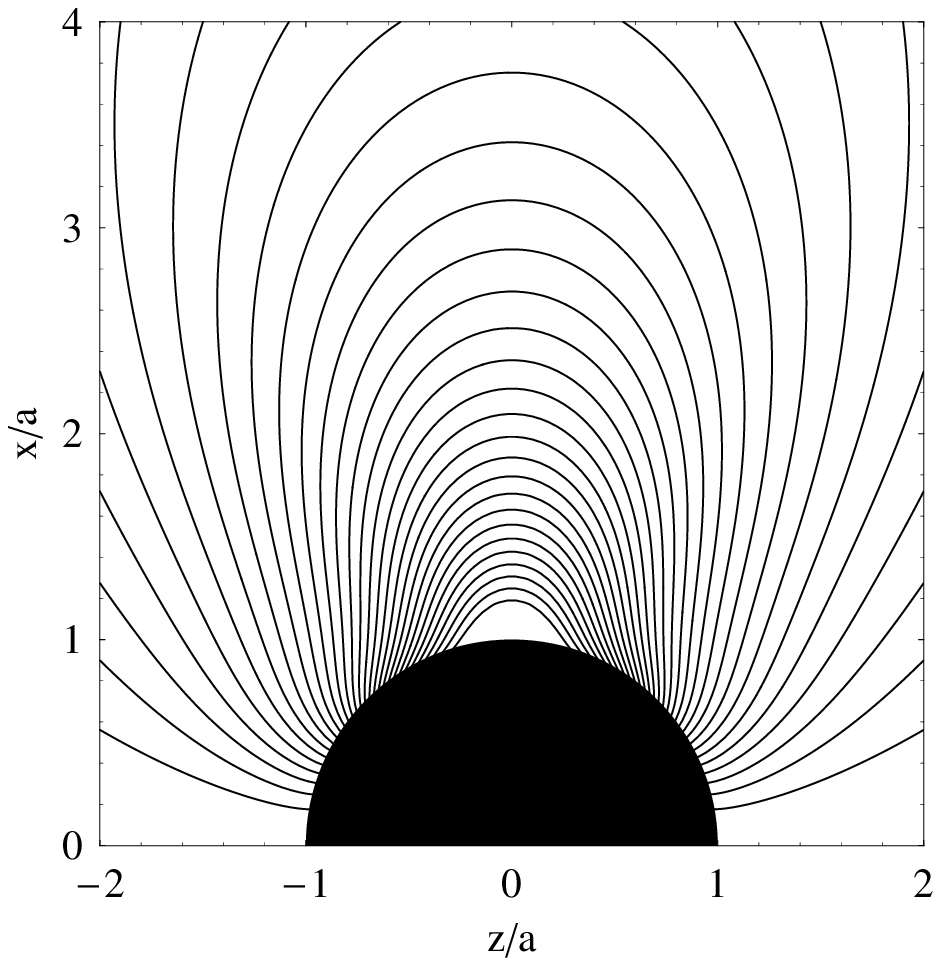}
   \put(-9.1,3.1){}
\put(-1.2,-.2){}
  \caption{}
\end{figure}
\newpage
\clearpage
\newpage
\setlength{\unitlength}{1cm}
\begin{figure}
 \includegraphics{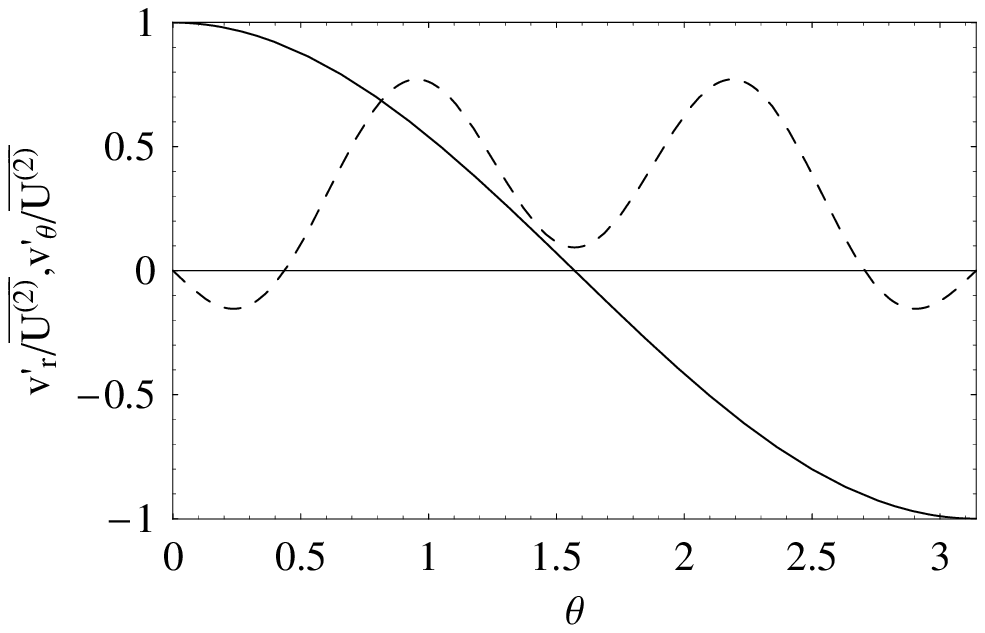}
   \put(-9.1,3.1){}
\put(-1.2,-.2){}
  \caption{}
\end{figure}
\newpage
\clearpage
\newpage
\setlength{\unitlength}{1cm}
\begin{figure}
 \includegraphics{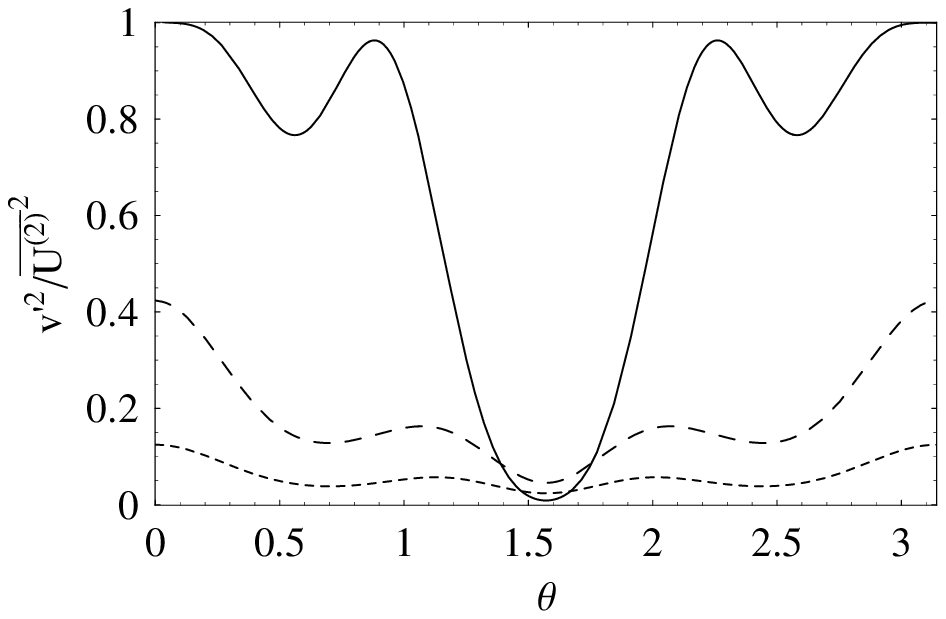}
   \put(-9.1,3.1){}
\put(-1.2,-.2){}
  \caption{}
\end{figure}
\newpage
\clearpage

\end{document}